\documentclass[slac_one]{revtex4}
\usepackage{graphicx}
\usepackage{fancyhdr}
\begin{document}

\title{Report from the IUPAP Commission on Particles and Fields (C11)} 

%

\author{Gregor Herten}
\affiliation{University of Freiburg, 79104 Freiburg, Germany}
%

\begin{abstract}
The Commission 11 (C11) on Particles and Fields of the International Union of Pure and Applied Physics (IUPAP) 
supervises the organization of the main conferences in particle physics. In the period 2007-2008 it has been active in the oversight and venue selection for the ICHEP and Lepton Photon Conferences, establishment of the new conference on Technology and Instrumentation in Particle Physics (TIPP), the selection of the prize winners for the IUPAP Young Scientist Prize in Particle Physics and the establishment of a working group on  the "Assessment of Individual Achievements in Large Collaborations in Particle Physics".  
\end{abstract}

\maketitle

\thispagestyle{fancy}


\section{The IUPAP Commission on Particles and Fields} 
 The International Council for Science (ICSU) \footnote{http://www.icsu.org} acts as an umbrella organization for  
various fields of science.  It has 29 scientific union members representing fields of science and 114 national scientific members. The union for physics is the International Union of Pure and Applied Physics (IUPAP)\footnote{http://www.iupap.org}. It was established in 1922. Currently it has 59 national scientific members, as well as 19 commissions and 7 working groups  for various fields of physics, among them the Commission 11 on Particles and Fields (C11)\footnote{http://www.iupap.org/commissions/iupap-commissions.html}. The mandate of C11 is to promote the exchange of information in the field,  supervise the organization of the major conferences in particle physics, promote the free circulation of scientists, and maintain the liaison between IUPAP and the  International Committee for Future Accelerators (ICFA)~\footnote{http://www.fnal.gov/directorate/icfa/}, which was established by C11 in 1976.
At the time of the ICHEP-08 Conference C11 has 13 members and 2 associated members from other commissions \cite{c11}. 

The report summarizes 
the main activities of C11 in the period 2007-2008
\section{Sponsored conferences} 
C11 has a tradition of more than 40 years to sponsor and supervise the main conferences in particle physics.  These are the International Conference on High Energy Physics (ICHEP) and the Symposium on Lepton-Photon Interactions (LP). In the past years these conferences have been held in Moscow (ICHEP-06), Daegu/Korea (LP-07) and Philadelphia (ICHEP-08).  C11 has decided that the future conferences will be the LP-09 in Hamburg, ICHEP-10 in Paris, LP-11 in Mumbai and ICHEP-12 in Melbourne. 

Besides the two main conferences C11 has started recently to sponsor additional conferences in particle physics. The first was the  Neutrino 08 Conference in Christchurch, New Zealand. In additional C11 took the initiative to establish a new conference on Technology and Instrumentation in Particle Physics (TIPP). The first conference of this type (TIPP09) will be held in Tsukuba, Japan, in March 2009. With parallel and plenary sessions it will be similar in style to the ICHEP conference series.

\section{IUPAP Young Scientist Prize}
In 2005 IUPAP has established a young scientist prize for all Commissions. The candidates should have a maximum  of 8 years of research after the Ph.D. to qualify for the prize.    The first prize in particle physics has been awarded in 2008 at the ICHEP-08 conference. C11 has asked for nominations by letters to all particle physics institutes world wide  and information provided by electronic mail through national contact physicists . In total 24 nominations have been received. The C11 commission members acted as the prize committee, receiving support from many external referees who were acquainted with the scientific work of the nominated candidates. The prize committee  decided to award the IUPAP Young Scientist Prize 2008 to Kai-Feng Chen, Taipei, in experimental and Yasaman Farzan, Tehran, in theoretical particle physics. The prize consists of a medal, a certificate and a small monetary award.  

Kai-Feng Chen receives the prize in recognition of his innovative contributions to the analysis of B-meson decays with the Belle experiment, among them time-dependent CP violation and polarization measurements in B-decays. 

Yasaman Farzan receives the Prize in recognition of her innovative theoretical contributions to neutrino and lepton physics, among them CP-violation, electric dipole moments of leptons and supernova cooling processes involving new weakly interacting particles.

 \section{Assessment of individual achievements in large collaborations}

C11 has set up a working group \cite{c11wg}, which has issued a report about the assessment of individual achievements in large collaborations. The report is accessible on the IUPAP/C11 web page. The conclusions are summarized here.     

Experiments in elementary particle physics  are only possible in a world-wide collaboration of many scientists. The detectors are designed, constructed and operated by large research groups and the scientific results are a common achievement of many scientists. The time required from the first idea about the experiment, design, construction to data taking and data analysis spans typically more than ten or twenty years. During that period, the continuous contributions of all participants Ð experts in detector hardware, calibration and data analysis alike Ð are essential for publishing scientific results. Therefore, it is customary in experimental particle physics for publications to be signed by many authors in alphabetic order.  

This procedure implies that an assessment of scientific achievements based mainly on publication lists and impact factors is no longer applicable in experimental particle physics. More factors must be included to judge the scientific merits of individual researchers in this field. 

For these reasons IUPAP/C11 has set up the working group to address these issues. 
This group included representatives of large collaborations in Particle Physics. Through these representatives, the collaborations were invited to comment on a draft version of this report.

\subsection{Goals}

The goal of the exercise was to define a common set of measures to enhance the visibility of individual achievements while maintaining the successful collaborative spirit in large collaborations in particle physics.

A common catalogue of objective criteria should be established, which should help to assess individual achievements. 

The criteria should be transparent to decision makers outside of the large collaborations, such as at universities, laboratories and prize committees. 

The Commission C11 encourages the collaborations in particle physics to agree to a common set of criteria and measures and to adapt their internal procedures accordingly while recognizing that the final decision rests with each collaboration.  

Decision makers in selection, promotion and prize committees at universities and science organizations should use these established criteria to assess the achievements of particle physicists and compare them to scientists in other fields. 

This catalogue of criteria could be used in other fields of science, where large collaborations are required to achieve results.

\subsection{Proposed measures} 

The working group proposes the following measures:
\begin{itemize}
\item{Eligible authors\newline
The Collaborations shall have clear internal rules regarding who is an eligible author for each publication. The rules shall be public and transparent and follow the guideline that Òauthorship should be limited to those who have made a significant contribution to the concept, design, execution and interpretation of the research study. All those who have made significant contributions should be offered the opportunity to be listed as authors.Ó (See, as an example: American Physical Society Guidelines for Professional Conduct)\cite{aps}.}    
\item{Publication Web page\newline
For each publication the collaboration might release a public web page with supporting notes and details about the individual contributions in analysis, operation, calibration, computing, editorial, etc., which have been essential for the publication. }
\item{Most relevant publications\newline
Rather than a list of all publications, one finds often in the curriculum vitae of experimental particle physicists a list of Òmost relevant publications.Ó This could be an indicator for scientific merit if the criteria for Òmost relevant publicationsÓ are clearly defined and similar in all collaborations. A good criterion for including a publication in this list could be the appearance of the individual as a significant contributor on the publication web page of the collaboration. 
}
\item{Scientific and technical notes\newline 
Scientific and technical notes, published by a few authors in an internally or externally refereed form, could help to make individual contributions more visible. These notes can describe in more detail the detector development, operation and calibration, as well as reconstruction algorithms, analysis techniques and statistical methods. 
}
\item{Public track record\newline
Collaborations should keep a public track record of authorship of refereed internal notes (listed with author names and titles of the notes), nominated speakers for conferences, responsibilities and positions inside the collaboration (with an explanation about the scientific merits required for this task), contributions to the operation of the experiment, membership in editorial boards, and other positions of responsibility.
}
\item{Two-tier author list\newline
Collaborations could consider the use of a two-tiered author list to emphasize special contributions to publications.  One option is to list a group of Òmain authorsÓ, another option is to keep the alphabetical order but mark some names as principal authors. }
\item{Awards\newline
Awards are an important measure to make individual achievements in large collaborations known to outside people. More use should be made of awards in particle physics: HEP-wide prizes, awards in countries, laboratories and universities as well as inside collaborations to acknowledge the scientific achievements of scientists (e.g., for PhD theses, data analysis, detector development, detector operation and calibration). 
}
\end{itemize}
\begin{acknowledgments} 
The author 
wishes to thank  the  members of the C11 commission and the working group  for their support.    
\end{acknowledgments}â

\end{document}